\documentclass[conference]{IEEEtran}
\ifCLASSINFOpdf
\else
\fi
\usepackage{amssymb,amsmath,epsfig,tabularx,delarray,enumerate,subfigure,graphicx,array,cite}

\begin{document}
%
\title{Double-Directional Information Azimuth Spectrum and Relay Network Tomography for a Decentralized Wireless Relay Network}

\author{\IEEEauthorblockN{Yifan Chen}
\IEEEauthorblockA{School of Engineering\\
University of Greenwich\\
Chatham Maritime, Kent ME4 4TB\\
Email: Y.Chen@gre.ac.uk} \and \IEEEauthorblockN{Chau Yuen}
\IEEEauthorblockA{Institute for Infocomm Research\\
1 Fusionopolis Way, 21-01 Connexis\\
Singapore 138632\\
Email: cyuen@i2r.a-star.edu.sg}}


%


\maketitle

\begin{abstract}
A novel channel representation for a two-hop decentralized wireless
relay network (DWRN) is proposed, where the relays operate in a
completely distributive fashion. The modeling paradigm applies an
analogous approach to the description method for a
double-directional multipath propagation channel, and takes into
account the finite system spatial resolution and the extended relay
listening/transmitting time. Specifically, the double-directional
information azimuth spectrum (IAS) is formulated to provide a
compact representation of information flows in a DWRN. The proposed
channel representation is then analyzed from a geometrically-based
statistical modeling perspective. Finally, we look into the problem
of relay network tomography (RNT), which solves an inverse problem
to infer the internal structure of a DWRN by using the instantaneous
double-directional IAS recorded at multiple measuring nodes exterior
to the relay region.
\end{abstract}


%

\section{Introduction}
%
%
%
%
\PARstart{R}ecently a great deal of research has been devoted to
cooperative wireless communications \cite{LTW04,WZZ06,SS07}, which
reap spatial diversity benefits from a virtual antenna array formed
with multiple relay nodes. In a cooperative scheme, one user
terminal (UT) partners with another UT to send its signal to the
access point (AP) or some other final destinations. The partner UT
serves as a relay, forwarding the message from the source to the
destination. For most of the existing works, implementation of the
schemes assumes central coordination among relays. Nevertheless, the
overhead to set up the cooperation may drastically reduce the useful
throughput \cite{OLT07}. Furthermore, a centralized operation may
not be feasible for a distributed system with a dynamic
infrastructure (e.g., a wireless \emph{ad hoc} network) or with a
energy-constrained operational mode (e.g., a wireless sensor
network). In this paper, we consider a decentralized wireless relay
network (DWRN) without link combining and joint scheduling among
relays \cite{CHS08}.

In our previous work \cite{CR09}, the performance of a DWRN was
analyzed by realizing the noteworthy analogy between a virtual DWRN
and a physical propagation channel. The information delivered from
the source to the destination flows through a DWRN just as the
signal power flows through a physical channel. The concept of
information azimuth-delay spectrum (IADS) was defined, which is
parallel to the power azimuth-delay spectrum (PADS) used in
describing a \emph{single-directional} physical channel
\cite{PMF00,MOL05}. However, the methodology in \cite{CR09} is only
applicable to a perfect receiver, which has an infinite
angle-resolving capability and incurs zero data processing delay.
Furthermore, the relay is supposed to listen and forward the data
within very short intervals, comparable to the propagation delays of
wireless signals in a DWRN. The current work revisits the analysis
in \cite{CR09} by looking into more general operation scenarios with
finite system sensitivity and extended relay listening and
transmitting periods. This can be achieved by using a
\emph{double-directional} description of the relay network, parallel
to the double-directional characterization of the propagation
channel \cite{MOL05}. Moreover, the DWRN is sampled at
uniformly-spaced angles-of-departure (AODs) and angles-of-arrival
(AOAs), leading to a \emph{discrete} information azimuth spectrum
(IAS). The problem of interest is thus to determine the parameters
of the discrete IAS to achieve both accuracy with respect to the
continuous benchmark model and coherence with respect to the
resolution limit of the systems. Subsequently, the novel channel
representation will be analyzed from a geometrically-based
statistical modeling perspective \cite{ER99,CR09}, which assumes
certain geometric distributions of relay nodes and then derives the
IAS by applying the fundamental information theory.

Provided with the network description preliminaries, the concept of
relay network tomography (RNT) is introduced, which is applied to
identify relay locations in an unknown DWRN. Probing signals are
transmitted from several exterior sources to illuminate the service
region, and relayed signals at several exterior destinations are
measured. As a result, the spatial distribution of active relays is
obtained from the received instantaneous IAS by solving an inverse
problem, which provides critical insight into the performance
bottlenecks in a network. Apparently, the basic principle of RNT
bears a strong resemblance to other inverse problems, in which key
aspects of a system are not directly observable. The term tomography
is coined to link the problem of interest here, in concept, to other
processes that infer the internal characteristics of an object from
external observation, as done in medical tomography \cite{KS01}.

The rest of the paper is organized as follows. In Section II, we
introduce the system model of a DWRN with realistic operation
scenarios and formulate several principal quantities for description
of such networks. Subsequently, the proposed channel representation
under the framework of geometric network modeling is investigated in
Section III, for which the relevant channel quantities are derived.
In Section IV, we provide a general formulation of the RNT problem.
Section V demonstrates the properties of the model parameters and
the efficacy of the RNT methodology through several numerical
examples. Finally, some concluding remarks are drawn in Section VI.

\section{Double-Directional Description of DWRN}
\subsection{System Model}
We will consider a DWRN with $L+2$ nodes: A source $\mathbf{S}$, a
destination $\mathbf{D}$, and $L$ non-collocated relays
$\mathbf{R}_1,\mathbf{R}_2,\cdots,\mathbf{R}_L$ as illustrated in
Fig. 1. The relays are randomly scattered in a given region
$\mathcal{R}$, which are due to the irregular nature of wireless
sensor or \emph{ad hoc} networks. In a sensor network, a large
number of relays with sensing capabilities may be spread over the
site under investigation in a distributive and randomized manner. On
the other hand, each UT in an \emph{ad hoc} network can serve as an
assisting relay and the non-stationary network architecture may
cause uncertainty to the relay positions \cite{AJH08}. Subsequently,
the following assumptions on the transmission strategy are imposed.
\begin{enumerate}
    \item \emph{Relay Distribution:} The locations of $\mathbf{R}_l~(l=1,2,\cdots,L)$ are described by
    a statistical density function $f_{\mathbf{R}}\left(\mathbf{R}_l\right)~(\mathbf{R}_l\in\mathcal{R})$, which represents the ensemble
    of a large number of randomly distributed relays.
    \item \emph{Relay Selection and Cooperation:} The network geometry is unknown to either $\mathbf{S}$ or
    $\mathbf{D}$. Thus, $\mathbf{S}$ broadcasts the same message to all $\mathbf{R}_l$
    and coherent relaying is not employed (i.e., without link combining and joint scheduling among relays).
    \begin{figure} [!htp]
\begin{center}
\epsfig{file=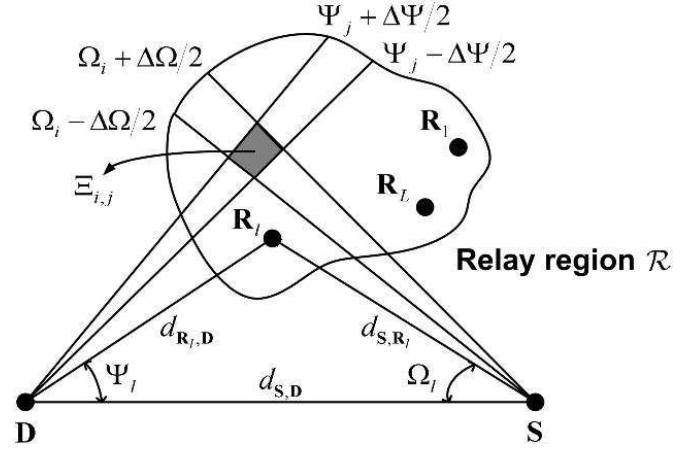,width=\linewidth}
\end{center}
    \caption{Double-directional DWRN channel}
    \label{fig:Fig1}
\end{figure}\item \emph{Decode-and-forward Processing:} Each $\mathbf{R}_l$ receives information from $\mathbf{S}$ in the first hop (backward channel)
    and then forwards the decoded signal to $\mathbf{D}$ in the second hop (forward channel). Two orthogonal frequency tones
    are used at these two transmission phases, respectively.
    \item \emph{Medium Access:} Due to the decentralized and unregulated nature of relay positions, a simplified time-division protocol is
    applied, which requires no knowledge of the network geometry at either $\mathbf{S}$ or $\mathbf{D}$. At the
    initial time $t=t_0$, $\mathbf{S}$ transmits its signal to all $\mathbf{R}_l$ $(l=1,2,\cdots,L)$. For the $l\mathrm{th}$ relay, it
    listens to the transmission from $\mathbf{S}$ in the time interval $\left[t_0,t_0+{\Delta}t_l^{(\mathbf{B})}\right]$.
    After processing the received signals from $\mathbf{S}$, $\mathbf{R}_l$ forwards the message to $\mathbf{D}$ within
    $\left[t_0+{\Delta}t_l^{(\mathbf{B})},t_0+{\Delta}t_l^{(\mathbf{B})}
    +{\Delta}t_l^{(\mathbf{F})}\right]$. The two parameters ${\Delta}t_l^{(\mathbf{B})}$ and
    ${\Delta}t_l^{(\mathbf{F})}$ denote the time spent in the
    backward and forward phases, respectively. The same time-division process repeats in the subsequent time slots
    $\Big[t_0+k\times\left({\Delta}t_l^{(\mathbf{B})}
    +{\Delta}t_l^{(\mathbf{F})}\right),t_0+(k+1)\times\left({\Delta}t_l^{(\mathbf{B})}
    +{\Delta}t_l^{(\mathbf{F})}\right)\Big]$
    $(k=1,2,\cdots)$.  For simplicity, each relay spends equal times in its listening and forwarding phases.
    Apparently, $\mathbf{R}_l$ introduces an AOD $\Omega_l$ and an AOA $\Psi_l$ during each transmission cycle as shown in Fig.
    1. It is worth emphasizing that the time-of-arrival (TOA)
    $\tau_l=\frac{d_{\mathbf{S},\mathbf{R}_l}+d_{\mathbf{R}_l,\mathbf{D}}}{c}$,
    which is one of the key quantities used in the
    multipath channels \cite{MOL05}, has lost its physical meaning in
    the current context. Here $d_{\mathbf{S},\mathbf{R}_l}$ and
    $d_{\mathbf{R}_l,\mathbf{D}}$ are the distances of the links
    $\mathbf{S}\rightarrow\mathbf{R}_l$ and $\mathbf{R}_l\rightarrow\mathbf{D}$, respectively, and $c$ is the speed of electromagnetic waves.
    \item \emph{Frequency-Tone Assignment:} In general, mutual interference among the forward
    channels may incur due to the simultaneously transmitting relays. One way to eliminate the interference is to assign $L$ frequency tones drawn
    from the pool of available bandwidth $\left[\omega_{\mathrm{min}},\omega_\mathrm{max}\right]$ to $L$ relay-to-destination pairs, all of which are also orthogonal
    to the carrier allocated to the backward channels.
\end{enumerate}

\subsection{Double-Directional Representation of DWRN}
For a two-hop DWRN, multiple information pipelines are laid via the
$L$ links
$\mathbf{S}\rightarrow\mathbf{R}_1\rightarrow\mathbf{D},\cdots,\mathbf{S}\rightarrow\mathbf{R}_L\rightarrow\mathbf{D}$.
Furthermore, the intersection of each AOD and each AOA uniquely
determines the location of a relay. Therefore, as an ideal reference
model, the information flows in the network are continuously
distributed in the AOD-AOA domain as
\begin{equation} 
I_{\mathrm{D}}(\Omega,\Psi)=\sum^L_{l=1}I_l\delta\left(\Omega-\Omega_l\right)\delta\left(\Psi-\Psi_l\right)
\end{equation}
where $I_\mathrm{D}(\Omega,\Psi)$ is defined as the
double-directional IAS and $I_l$ is the outage capacity given an
outage probability of $P_\mathrm{out}$ for the link
$\mathbf{S}\rightarrow\mathbf{R}_l\rightarrow\mathbf{D}$. The
terminology double-directional is adopted according to \cite{MOL05}.

Eq. (1) assumes perfect receivers that are able to resolve the
departing and arriving information flows with infinite sensitivity
in the angular domain. Therefore, the formulation represents an
ideal benchmark model. In practice, both the source and destination
have limited system sensitivity. Consider two data streams via two
relays $\mathbf{R}_{l_1}$ and $\mathbf{R}_{l_2}$
$(1{\leq}l_1{\neq}l_2{\leq}L)$. Due to the finite length of the
transmit (receive) antenna aperture, the AODs (AOAs) of these two
information flows cannot be successfully distinguished if the
separation of their departing (impinging) angles,
$|\Omega_{l_1}-\Omega_{l_2}|$ $(|\Psi_{l_1}-\Psi_{l_2}|)$, is within
the resolution limit of the source (destination) antenna,
$\Delta\Omega$ $(\Delta\Psi)$. Therefore, the DWRN channel should be
sampled at uniformly-spaced AODs and AOAs:
\begin{equation} 
\begin{split}
\Omega_i=i\Delta\Omega,~~~~~~~~~~\left\lfloor\frac{\Omega_\mathrm{min}}{\Delta\Omega}\right\rfloor{\leq}i
{\leq}\left\lceil\frac{\Omega_\mathrm{max}}{\Delta\Omega}\right\rceil\\
\end{split}
\end{equation}
\begin{equation} 
\begin{split}
\Psi_j=j\Delta\Psi,~~~~~~~~~~\left\lfloor\frac{\Psi_\mathrm{min}}{\Delta\Psi}\right\rfloor{\leq}j
{\leq}\left\lceil\frac{\Psi_\mathrm{max}}{\Delta\Psi}\right\rceil\\
\end{split}
\end{equation}
where ${\lfloor}\cdot\rfloor$ and ${\lceil}\cdot\rceil$ are the
floor and ceiling functions, respectively.

Subsequently, the discrete double-directional IAS can be expressed
as
\begin{equation} 
\begin{split}
I^{(\mathrm{d})}_{\mathrm{D}}\left(\Omega_i,\Psi_j\right)&=\iint_{\Xi_{i,j}}I_{\mathrm{D}}(\Omega,\Psi)f^{(i,j)}_{\Omega,\Psi}
(\Omega,\Psi)d{\Omega}d\Psi,\\
\mathrm{where}~\Xi_{i,j}&\triangleq\Bigg\{\left(\Omega,\Psi\right):\Omega_i-\frac{\Delta\Omega}{2}\leq\Omega\leq\Omega_i+\frac{\Delta\Omega}{2}\\
&~~~~~~~~~~~~~\&~\Psi_j-\frac{\Delta\Psi}{2}\leq\Psi\leq\Psi_j+\frac{\Delta\Psi}{2}\Bigg\}
\end{split}
\end{equation}
In (4), The superscript $^{(\mathrm{d})}$ denotes a discrete model.
The \emph{local} joint AOD-AOA pdf
$f^{(i,j)}_{\Omega,\Psi}(\Omega,\Psi)$ satisfies
$\iint_{\Xi_{i,j}}f^{(i,j)}_{\Omega,\Psi}(\Omega,\Psi)d{\Omega}d\Psi=1$,
where the integration domain $\Xi_{i,j}$ is illustrated in Fig. 1.

\section{Geometrically-Based Statistical Model for a Random DWRN}
We will discuss the proposed analytical framework in Section II from
a geometrically-based statistical modeling perspective \cite{CR09},
where a large number of relays are randomly located in the
two-dimensional space according to a specified relay density
function. This approach is useful when the spatial structure
observed in a large DWRN is far from being regular \cite{AJH08}. The
medium access arrangement follows the protocol in Section
II-\emph{A}. Moreover, all channels experience independent
frequency-flat block fading with the complex channel gain between
the link $\mathbf{M}\rightarrow\mathbf{N}$,
$h_{\mathbf{M},\mathbf{N}}$
$\left(\mathbf{M}\in\left\{\mathbf{S},\mathbf{R}_1,\cdots,\mathbf{R}_L\right\}
,\mathbf{N}\in\left\{\mathbf{D},\mathbf{R}_1,\cdots,\mathbf{R}_L\right\}\right)$
following the Nakagami distribution with shape parameter $m$
\cite{MOL05}. \begin{figure} [!htp]
\begin{center}
\epsfig{file=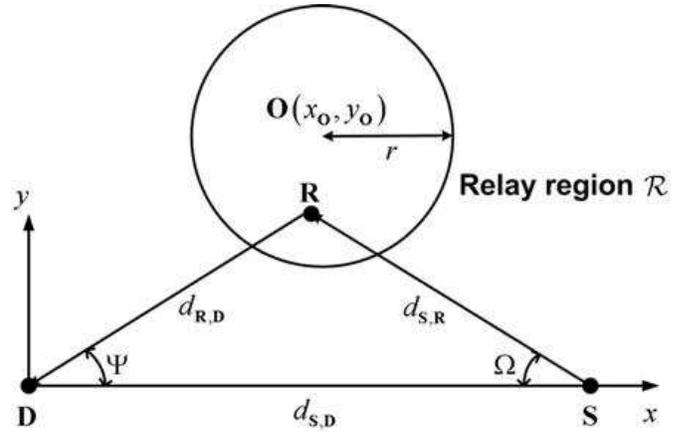,width=\linewidth}
\end{center}
    \caption{Geometrically-based statistical model for a DWRN}
    \label{fig:Fig2}
\end{figure}The associated instantaneous power
$|h_{\mathbf{M},\mathbf{N}}|^2$ is gamma-distributed with the same
shape parameter, i.e.,
$|h_{\mathbf{M},\mathbf{N}}|^2\sim\mathcal{G}\left(m,\frac{\lambda_{\mathbf{M},\mathbf{N}}}{m}\right)$,
where $\mathcal{G}(\kappa,\theta)$ represents the gamma distribution
with scale $\theta$ and shape $\kappa$. The mean value
$\lambda_{\mathbf{M},\mathbf{N}}=d^\nu_{\mathbf{M},\mathbf{N}}$ with
$\nu$ being the path loss exponent. The ambient noise at the relays
and the destination, $Z_{\mathbf{N}}\sim\mathcal{N}(0,N_0)$, where
$\mathcal{N}(\mu,\sigma^2)$ represents a normal distribution with
mean $\mu$ and variance $\sigma^2$. The communication bandwidth is
$B$ for any channel. Subsequently, the average received
signal-to-noise ratio (SNR) for each link is
$\mathrm{SNR}=\frac{P}{BN_0}$, where $P$ is the average transmit
power for all transmitting terminals. Finally, the channel state
information is assumed to be known at the receivers but unavailable
to the transmitters \cite{LTW04}.

Let the $x$-$y$ coordinate system be defined such that the
destination is at the origin and the source lies on the $x$ axis, as
shown in Fig. 2. It is assumed that there are many relay nodes, the
locations of which are described by the statistical relay density
function $f_{x,y}(x,y)$. By applying the law of sines to the
triangle $\mathbf{SRD}$ in Fig. 2, the relay-to-destination distance
$d_{\mathbf{R},\mathbf{D}}$ can be expressed in terms of the AOD
$\Omega$ and the AOA $\Psi$ as
\begin{equation} 
d_{\mathbf{R},\mathbf{D}}=\frac{d_{\mathbf{S},\mathbf{D}}\sin\Omega}{\sin(\Omega+\Psi)}
\end{equation}
The source-to-relay distance $d_{\mathbf{S},\mathbf{R}}$ is
similarly derived as
\begin{equation} 
d_{\mathbf{S},\mathbf{R}}=\frac{d_{\mathbf{S},\mathbf{D}}\sin\Psi}{\sin(\Omega+\Psi)}
\end{equation}

\emph{Proposition 1 (Derivation of $f_{\Omega,\Psi}(\Omega,\Psi)$):}

The joint AOD-AOA pdf of information flows within a DWRN is given by
\begin{equation} 
\begin{split}
f_{\Omega,\Psi}(\Omega,\Psi)&=\frac{d^2_{\mathbf{S},\mathbf{D}}\left|\sin2\Omega+\sin2\Psi-\sin2(\Omega+\Psi)\right|}
{\left[1-\cos2(\Omega+\Psi)\right]^2}\\
&{\times}f_{x,y}\left(\frac{d_{\mathbf{\mathbf{S},\mathbf{D}}}\sin\Omega\cos\Psi}{\sin(\Omega+\Psi)},
\frac{d_{\mathbf{\mathbf{S},\mathbf{D}}}\sin\Omega\sin\Psi}{\sin(\Omega+\Psi)}\right)
\end{split}
\end{equation}
where $d_{\mathbf{S},\mathbf{D}}$ is the distance between the source
and the destination.

\emph{Proof:} The proof is omitted here for simplicity.

\emph{Proposition 2 (Derivation of $I_{\mathrm{D}}(\Omega,\Psi)$):}

The double-directional IAS $I_{\mathrm{D}}(\Omega,\Psi)$ is the
argument of the function
\begin{equation} 
\begin{split}
&\mathcal{P}\left(I\right)\\
&=1-\left[1-\frac{\gamma\left(m,\frac{m\left(4^{I}-1\right)}{\mathrm{SNR}{\times}d^\nu_{\mathbf{S},\mathbf{R}}}\right)}{\Gamma(m)}\right]
\left[1-\frac{\gamma\left(m,\frac{m\left(4^{I}-1\right)}{\mathrm{SNR}{\times}d^\nu_{\mathbf{R},\mathbf{D}}}\right)}{\Gamma(m)}\right]
\end{split}
\end{equation}
when $\mathcal{P}\left(I\right)=P_\mathrm{out}$. In (8),
$\gamma(\alpha,\beta)=\int^\beta_0t^{\alpha-1}e^{-t}dt$ is the lower
incomplete gamma function and $\Gamma(\cdot)$ is the gamma function.

\emph{Proof:} The proof is omitted here for simplicity.

Finally, the local joint AOD-AOA pdf in (4),
$f^{(i,j)}_{\Omega,\Psi}(\Omega,\Psi)$, is related to
$f_{\Omega,\Psi}(\Omega,\Psi)$ through the following relationship:
\begin{equation} 
f^{(i,j)}_{\Omega,\Psi}(\Omega,\Psi)=\frac{f_{\Omega,\Psi}(\Omega,\Psi)}{\iint_{\Xi_{i,j}}f_{\Omega,\Psi}(\Omega,\Psi)d{\Omega}d\Psi}
\end{equation}
Substituting (7)-(9) into (4) gives the discrete double-directional
IAS.

\section{RNT for a DWRN}
\subsection{Formulation of RNT}
In the general RNT problem depicted in Fig. 3, $L$ relays
$\left(\mathbf{R}_1,\cdots,\mathbf{R}_L\right)$ are randomly
scattered in a finite area $\mathcal{R}$ and $Q$ measuring nodes
$\left(\mathbf{N}_1,\cdots,\mathbf{N}_Q\right)$ are placed exterior
to $\mathcal{R}$. Nevertheless, the following analysis is also
applicable to any other measurement network orientation with respect
to $\mathcal{R}$. Each measuring node can serve as the probing
source as well as the information destination, which has a
transmit/receive azimuthal resolution of $\Delta\Theta$. As shown in
Fig. 3, the first node $\mathbf{N}_1$ transmits over its full
spectrum of AODs
$i\Delta\Theta~\left(i=\left\lfloor\frac{\Omega^{(1)}_{\min}}{\Delta\Theta}\right\rfloor,
\left\lfloor\frac{\Omega^{(1)}_{\min}}{\Delta\Theta}\right\rfloor+1,\cdots,
\left\lceil\frac{\Omega^{(1)}_{\max}}{\Delta\Theta}\right\rceil\right)$,
where the superscript $^{(1)}$ denotes $\mathbf{N}_1$. All other
nodes $\mathbf{N}_2\sim\mathbf{N}_Q$ record an information flow.
Each receive node $\mathbf{N}_q~(q=2,\cdots,Q)$ scans its entire
angular range
$j\Delta\Theta~\left(j=\left\lfloor\frac{\Psi^{(q)}_{\min}}{\Delta\Theta}\right\rfloor,
\left\lfloor\frac{\Psi^{(q)}_{\min}}{\Delta\Theta}\right\rfloor+1,\cdots,
\left\lceil\frac{\Psi^{(q)}_{\max}}{\Delta\Theta}\right\rceil\right)$,
where the superscript $^{(q)}$ denotes $\mathbf{N}_q$. At each
$j\Delta\Theta$, $\mathbf{N}_q$ searches the whole frequency band
within which the relays are operating. When $\mathbf{N}_q$ locks on
the $l{\mathrm{th}}$ frequency component $\omega_{l}$, which yields
a detectable information flow associated with the relay
$\mathbf{R}_l$, it then monitors the instantaneous channel capacity
over an observation period and records the time-variant capacity
$\hat{I}\left(t\right)$. From the set of $\hat{I}\left(t\right)$,
the channel outage capacity $\hat{I}^{(1,q)}_{l}$ can be estimated
empirically, where the superscript $^{(1,q)}$ denotes the
information pathway from $\mathbf{N}_1$ to $\mathbf{N}_q$. The
similar process continues until all the active relays are
identified. Then the second node $\mathbf{N}_2$ sends the probing
message and all the other nodes
$\mathbf{N}_1,\mathbf{N}_3,\cdots,\mathbf{N}_Q$ receive the data,
and so on.
\begin{figure} [!htp]
\begin{center}
\epsfig{file=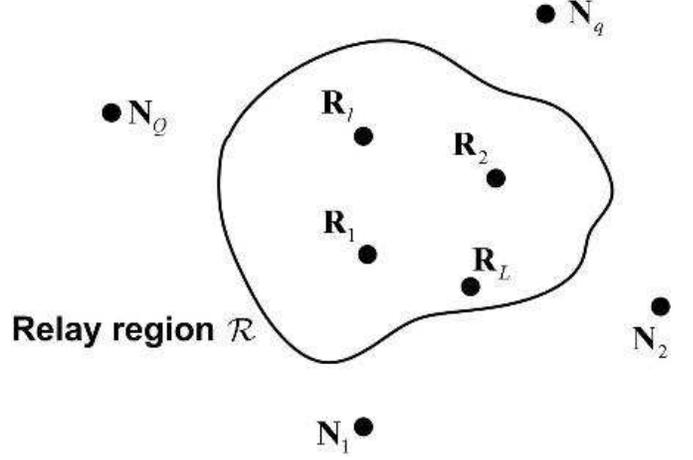,width=\linewidth}
\end{center}
    \caption{Formulation of RNT}
    \label{fig:Fig3}
\end{figure}

A procedure to calculate the double-directional IAS for a DWRN with
given sources, relays and destinations is called the \emph{direct
problem}. A procedure to obtain relay distribution knowing
sources/destinations and measuring the information flows at multiple
receivers is called the \emph{inverse problem}. The direct problem
has already been discussed in Section III. The corresponding inverse
problem can be formulated as follows.

\textbf{RNT Problem}: Given the set of measuring nodes
$\mathbf{N}_1,\cdots,\mathbf{N}_Q$, the estimated $(Q-1)^2{\times}L$
single-directional AOA matrix
\begin{equation} 
\begin{split}
\mathbf{\bar{\hat{\Psi}}}&=\left[
                                         \begin{array}{cccc}
                                         \hat{\mathbf{\Psi}}_1
                                         &
                                         \hat{\mathbf{\Psi}}_2
                                         &
                                         \cdots
                                         &
                                         \hat{\mathbf{\Psi}}_L\\
                                         \end{array}
                                       \right]\\
                                       &=
\left[
    \begin{array}{ccccc}
          \hat{\Psi}^{(1,2)}_1 & \hat{\Psi}^{(1,2)}_2 & \cdots & \hat{\Psi}^{(1,2)}_L \\
          \vdots & \vdots & \ddots & \vdots \\
          \hat{\Psi}^{(Q,Q-1)}_1 & \hat{\Psi}^{(Q,Q-1)}_2 & \cdots & \hat{\Psi}^{(Q,Q-1)}_L \\
    \end{array}
    \right]
    \end{split}
\end{equation}
and the estimated $(Q-1)^2{\times}L$ outage capacity matrix
\begin{equation} 
\begin{split}
\mathbf{\bar{\hat{I}}}&=\left[
                                         \begin{array}{cccc}
                                         \hat{\mathbf{I}}_1
                                         &
                                         \hat{\mathbf{I}}_2
                                         &
                                         \cdots
                                         &
                                         \hat{\mathbf{I}}_L\\
                                         \end{array}
                                       \right]\\
                                       &=
\left[
    \begin{array}{ccccc}
          \hat{I}^{(1,2)}_1 & \hat{I}^{(1,2)}_2 & \cdots & \hat{I}^{(1,2)}_L \\
          \vdots & \vdots & \ddots & \vdots \\
          \hat{I}^{(Q,Q-1)}_1 & \hat{I}^{(Q,Q-1)}_2 & \cdots & \hat{I}^{(Q,Q-1)}_L \\
    \end{array}
    \right]
    \end{split}
\end{equation}
$\hat{\Psi}^{\left(q_1,q_2\right)}_l$ and
$\hat{I}^{\left(q_1,q_2\right)}_l$ are the empirical AOA and channel
outage capacity associated with the information pipeline
$\mathbf{N}_{q_1}\rightarrow\mathbf{R}_l\rightarrow\mathbf{N}_{q_2}$,
respectively. Note that due to the channel reciprocity in a two-hop
DWRN, it is expected that
$\hat{\Psi}^{\left(q_1,q_2\right)}_l=\hat{\Omega}^{\left(q_2,q_1\right)}_l$
and
$\hat{I}^{\left(q_1,q_2\right)}_l=\hat{I}^{\left(q_2,q_1\right)}_l$.
Therefore, (11) is equivalent to a double-directional AOD-AOA
matrix, where half of the AOAs correspond to the AODs of the
reversed information pathways. Identify the locations of
$\mathbf{R}_1,\cdots,\mathbf{R}_L$ that satisfy the following two
constraint functions
\begin{equation} 
e_1\left(\mathbf{\hat{\Psi}}_l-\mathbf{\Psi}_l\right)\leq\epsilon_1,~~~~~~~~l=1,2,\cdots,L
\end{equation}
and
\begin{equation} 
e_2\left(\mathbf{\hat{I}}_l-\mathbf{I}_l\right)\leq\epsilon_2,~~~~~~~~l=1,2,\cdots,L
\end{equation}
$\mathbf{\Psi}_l$ and $\mathbf{I}_l$ are the AOA and channel outage
capacity vectors solved in the direct problem for all the
information paths via $\mathbf{R}_l$ assuming a specific location of
$\mathbf{R}_l$.

\subsection{Algorithms to Solve The RNT Problem}
In the current work, we will consider the $\ell^2$-norm of the two
error vectors, i.e.,
$e_1\left(\mathbf{\hat{\Psi}}_l-\mathbf{\Psi}_l\right)=\sqrt{\sum_{q_1}\sum_{q_2{\neq}q_1}
\left(\hat{\Psi}^{\left(q_1,q_2\right)}_l-\Psi^{\left(q_1,q_2\right)}_l\right)^2}$
and
$e_2\left(\mathbf{\hat{I}}_l-\mathbf{I}_l\right)=\sqrt{\sum_{q_1}\sum_{q_2{\neq}q_1}
\left(\hat{I}^{\left(q_1,q_2\right)}_l-I^{\left(q_1,q_2\right)}_l\right)^2}$.
The objectives are to minimize the values of $e_2$ and meanwhile
ensure that $e_1=0$ for all the detectable information flows. The
optimization process can be realized through the following steps:

$1)$ Discretize the solution domain of the relay
    locations $\mathcal{R}$ into $W$ sufficiently small square cells, i.e.,
    $\mathbb{R}=\mathcal{R}^{(\mathrm{d})}
    \triangleq\left\{\mathbf{x}_1,\cdots,\mathbf{x}_W\right\}$,
    where $\mathbf{x}_w=(x_w,y_w)~(w=1,2,\cdots,W)$ represents the Cartesian coordinates of the center of the $w{\mathrm{th}}$ cell;

$2)$ For each identified information pipeline
    $\mathbf{N}_{q_1}\rightarrow\mathbf{R}_l\rightarrow\mathbf{N}_{q_2}$,
    determine the subset of $\mathbb{R}$,
    $\mathbb{R}^{\left(q_1,q_2\right)}_l\subseteq\mathbb{R}$, such
    that
    ${\forall}\mathbf{x}\in\mathbb{R}^{\left(q_1,q_2\right)}_l$,
    $\hat{\Psi}^{\left(q_1,q_2\right)}_l=\Psi^{\left(q_1,q_2\right)}_l(\mathbf{x})$;

$3)$ Determine the joint of all the sets obtained in Step 2 as
$\mathbb{R}_l=\mathbb{R}^{(1,2)}_l\bigcap
    \mathbb{R}^{(1,3)}_l\bigcap
    \cdots\bigcap\mathbb{R}^{(Q,Q-1)}_l$;

$4)$ Estimate the location of $\mathbf{R}_l$ as
    \begin{equation} 
    \mathbf{x}\left(\mathbf{R}_l\right)=\mathrm{arg}\min_{\mathbf{x}\in\mathbb{R}_l}
    \sqrt{\sum_{q_1}\sum_{q_2{\neq}q_1}\left(\hat{I}^{\left(q_1,q_2\right)}_l
    -I^{\left(q_1,q_2\right)}_l(\mathbf{x})\right)^2}
    \end{equation}

It is not always possible to obtain an accurate estimate of the
outage capacity $\hat{I}^{\left(q_1,q_2\right)}_l$ due to the
limited observation data at each destination. In such a case, the
inverse problem could be approached from a hypothesis-testing
perspective. An alternative optimization procedure to Step 4 can be
formulated as follows.

$4^\star)$ Let there be $K$ relay-position hypotheses denoted
$\mathcal{H}_1:\mathbf{x}(\mathbf{R}_l)=\mathbf{\tilde{x}}_1$,$\mathcal{H}_2:\mathbf{x}(\mathbf{R}_l)=\mathbf{\tilde{x}}_2$,
$\cdots$,$\mathcal{H}_K:\mathbf{x}(\mathbf{R}_l)=\mathbf{\tilde{x}}_K$,
where $K=|\mathbb{R}_l|$ and
$\mathbb{R}_l=\left\{\mathbf{\tilde{x}}_1,\mathbf{\tilde{x}}_2,\cdots,\mathbf{\tilde{x}}_K\right\}$.
Furthermore, let $\varepsilon_{k_1,k_2}$ be the desired probability
of incorrectly selecting $\mathcal{H}_{k_2}$ given that
$\mathcal{H}_{k_1}$ is true. The sequential probability ratio test
procedures will be applied for statistical decision making, where
the decision at each stage is based on the likelihood ratio. In
particular, the multi-hypothesis sequential test is considered and
summarized below \cite{GVN07}. The likelihood ratio including prior
probabilities for a pair of hypotheses $k_1$ and $k_2$ after the
$o{\mathrm{th}}$ data observation is
\begin{equation} 
\eta^o_{k_1,k_2}=\frac{f_{\mathcal{H}_{k_1}}(\mathfrak{I}_1)f_{\mathcal{H}_{k_1}}(\mathfrak{I}_2){\cdots}f_{\mathcal{H}_{k_1}}(\mathfrak{I}_o)
\mathrm{Pr}\left(\mathcal{H}_{k_1}\right)}{f_{\mathcal{H}_{k_2}}(\mathfrak{I}_1)f_{\mathcal{H}_{k_2}}(\mathfrak{I}_2){\cdots}
f_{\mathcal{H}_{k_2}}(\mathfrak{I}_o)\mathrm{Pr}\left(\mathcal{H}_{k_2}\right)}
\end{equation}
where
$\mathfrak{I}_o=\left(\mathcal{I}^{(1,2)}_{l,o},\mathcal{I}^{(1,3)}_{l,o},\cdots,\mathcal{I}^{(Q,Q-1)}_{l,o}\right)^T$
is the instantaneous outage capacity vector at the $o{\mathrm{th}}$
observation for all the information routes via $\mathbf{R}_l$.
$f_{\mathcal{H}_{k_1}}(\mathfrak{I}_o)$ is the pdf of
$\mathfrak{I}_o$ under $\mathcal{H}_{k_1}$, which is given by
\begin{equation} 
\begin{split}
&f_{\mathcal{H}_{k_1}}(\mathfrak{I}_o)
\stackrel{(a)}{=}\prod_{q_1}\prod_{q_2{\neq}q_1}f_{\mathcal{I}}\left(\mathcal{I}^{(q_1,q_2)}_{l,o};\mathbf{x}(\mathbf{R}_l)
=\mathbf{\tilde{x}}_{k_1}\right)\\
&\stackrel{(b)}{=}\ln4\times\prod_{q_1}\prod_{q_2{\neq}q_1}\Bigg\{\left[\rho_1\left(\mathcal{I}^{(q_1,q_2)}_{l,o},\mathbf{\tilde{x}}_{k_1}\right)\right]^m
\\
&~~~~~~~~~~~~~~~~~~~~~~~~~~\times\exp\left[{-\rho_1\left(\mathcal{I}^{(q_1,q_2)}_{l,o},\mathbf{\tilde{x}}_{k_1}\right)}\right]\\
&~~~~~~~~~~~~~~~~~~~~~~~~~~\times\left[1-\frac{\gamma
\left(m,\rho_2\left(\mathcal{I}^{(q_1,q_2)}_{l,o},\mathbf{\tilde{x}}_{k_1}\right)\right)}{\Gamma(m)}\right]\\
&~~~~~~~~~~~~~~~~~~~~~~+\left[\rho_2\left(\mathcal{I}^{(q_1,q_2)}_{l,o}\right),\mathbf{\tilde{x}}_{k_1}\right]^m\\
&~~~~~~~~~~~~~~~~~~~~~~~~~~\times\exp\left[{-\rho_2\left(\mathcal{I}^{(q_1,q_2)}_{l,o},\mathbf{\tilde{x}}_{k_1}\right)}\right]\\
&~~~~~~~~~~~~~~~~~~~~~~~~~~\times\left[1-\frac{\gamma\left(m,\rho_1\left(\mathcal{I}^{(q_1,q_2)}_{l,o},\mathbf{\tilde{x}}_{k_1}\right)\right)}{\Gamma(m)}\right]\Bigg\}
\end{split}
\end{equation}
where $(a)$ follows the fact that the instantaneous outage
capacities are sequentially observed for all the information paths
via $\mathbf{R}_l$ during the measurement process. Therefore, they
can be assumed to be independently distributed. The equality in
$(b)$ is obtained by differentiating the cdf in (8) with respect to
the variable $I$. In (16),
\begin{equation}
\begin{nonumber}
\rho_1\left(\mathcal{I}^{(q_1,q_2)}_{l,o},\mathbf{\tilde{x}}_{k_1}\right)
=\frac{m\left(4^{\mathcal{I}^{(q_1,q_2)}_{l,o}}-1\right)}{\mathrm{SNR}{\times}\left[d_{\mathbf{S},\mathbf{R}}\left(\mathbf{\tilde{x}}_{k_1}\right)\right]^\nu}
\end{nonumber}
\end{equation}
and
\begin{equation}
\begin{nonumber}
\rho_2\left(\mathcal{I}^{(q_1,q_2)}_{l,o},\mathbf{\tilde{x}}_{k_1}\right)
=\frac{m\left(4^{\mathcal{I}^{(q_1,q_2)}_{l,o}}-1\right)}{\mathrm{SNR}{\times}\left[d_{\mathbf{R},\mathbf{D}}\left(\mathbf{\tilde{x}}_{k_1}\right)\right]^\nu}
\end{nonumber}
\end{equation}

If the number of observations is not fixed, a decision on the
relay-position state (i.e., $\mathcal{H}_{\hat{k}}$ is selected) is
made when the condition
\begin{equation} 
\eta^o_{\hat{k},k_2}>\frac{1-\varepsilon_{\hat{k},k_2}}{\varepsilon_{\hat{k},k_2}}~~~\forall\hat{k}{\neq}k_2
\end{equation}
is met. The threshold in (17) to which the likelihood ratio is
compared is designed to ensure that the average rate of making a
wrong decision in favor of $\mathcal{H}_{k_2}$ when
$\mathcal{H}_{k_1}$ is true is no more than $\varepsilon_{k_1,k_2}$.
Nevertheless, the inequality in (17) may not always be satisfied due
to the insufficient observation data. In such a case, a hypotheses
$\mathcal{H}_{\hat{k}}$ is selected if its associated likelihood has
the maximum value after a fixed number of $O$ observations, i.e.,
\begin{equation} 
\hat{k}=\mathrm{arg}\max_k~f_{\mathcal{H}_k}(\mathfrak{I}_1)f_{\mathcal{H}_k}(\mathfrak{I}_2){\cdots}f_{\mathcal{H}_k}(\mathfrak{I}_O)
\mathrm{Pr}\left(\mathcal{H}_k\right)
\end{equation}
Eqs. (17) and (18) provide alternative objective functions to the
one given in (14).
\begin{figure} [!htp]
\begin{center}
\epsfig{file=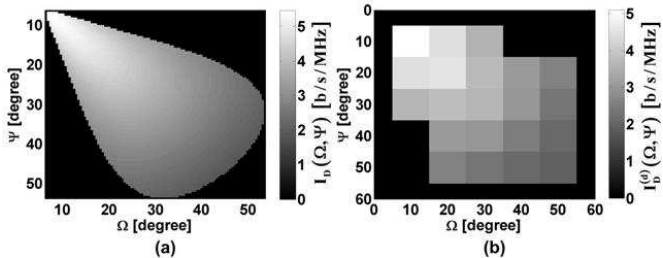,width=\linewidth}
\end{center}
    \caption{Double-directional IAS: (a) continuous reference model and (b) discrete model for systems with an angular resolution of $10^\circ$}
    \label{fig:Fig4}
\end{figure}
\begin{figure} [!htp]
\begin{center}
\epsfig{file=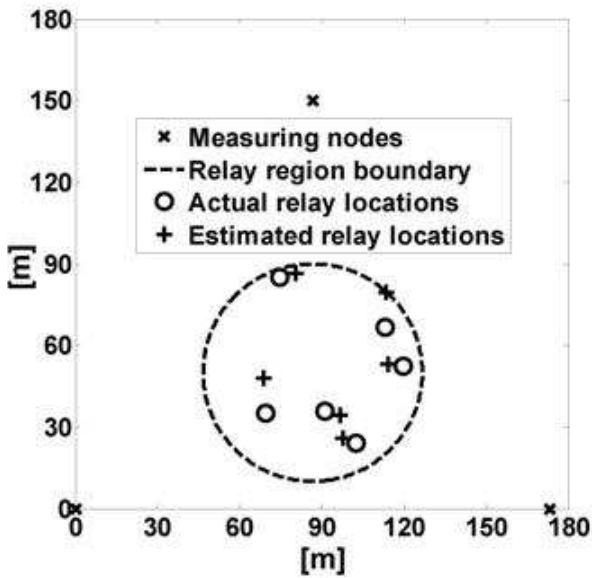,width=0.9\linewidth}
\end{center}
    \caption{Estimation of relay locations using the proposed RNT algorithm}
    \label{fig:Fig5}
\end{figure}
\section{Numerical Examples}
We consider the high SNR regime ($\mathrm{SNR}=30~\mathrm{dB}$) for
a fixed outage probability $P_\mathrm{out}=0.01$. The shape
parameter of the Nakagami distribution is $m=1$ (Rayleigh fading).
The distance between the source and the destination
$d_{\mathbf{S},\mathbf{D}}$ is assumed to be
$100\sqrt{3}~\mathrm{m}$ (see also Fig. 2) and the path loss
exponent is $\nu=-3$. The relay region is assumed to be a circle
centered at $\mathbf{O}(50\sqrt{3}~\mathrm{m},50~\mathrm{m})$ with a
radius $r=40~\mathrm{m}$. Finally, the spatial resolutions of both
the source and destination antennas are $10^\circ$.

Fig. 4 illustrates the double-directional IAS for both the
continuous reference model in (1) and its discrete counterpart in
(4). Apparently, both $I_\mathrm{D}(\Omega,\Psi)$ and
$I^{(\mathrm{d})}_\mathrm{D}(\Omega,\Psi)$ decrease as the AOD
$\Omega$ and the AOA $\Psi$ increase. This certainly holds an
intuitive appeal because larger values of $\Omega$ and $\Psi$ lead
to more severe path loss, thereby reducing the outage capacity of a
relay path. As a practical concern, Fig. 4 provides an intuitive and
quantitative answer to the following two important questions:
\emph{how will the information flows be scattered in the AOD and AOA
domains} and \emph{what will be the beam-scanning ranges of the
source and destination antennas in order to achieve sufficiently
high outage capacities?}

Next, we look into the RNT problem discussed in Section IV. A
three-node measurement network is deployed as illustrated in Fig. 5.
A number of relays are uniformly distributed within a circular
region. The procedures presented in Section IV-\emph{B} are applied
to estimate the relay locations, where the objective function of
(18) is used. The number of observations of the instantaneous
channel capacity for each information pipeline is $O=10$.
Furthermore, no \emph{a priori} knowledge of the probability of each
location under hypothesis test is available. As can be observed from
Fig. 5, most of the estimated relay locations agree well with the
actual ones, which demonstrates the efficacy of the proposed RNT
algorithm by using a relatively small number of measuring nodes and
observations.

\section{Conclusions}
We have presented a double-directional description method for
analyzing two-hop DWRN channels, which is based on an analogous
approach to the double-directional physical wireless channels. The
proposed model has been investigated using a geometrically-based
statistical modeling approach with uniformly distributed relay
nodes. We have also investigated the RNT problem by applying the
proposed network representation. Finally, numerical examples have
been used to demonstrate the practical significance of the proposed
analytical framework.



%

\bibliographystyle{IEEEbib}
\bibliography{IEEEbib}

\end{document}